\documentclass[12pt]{article}

\usepackage{amsmath,graphicx}
\usepackage{epsf}
\usepackage{graphicx,epsfig}
\usepackage{amsfonts}
\usepackage{amssymb}
\usepackage[nosort]{cite}
\usepackage{setspace}

\def\bk{{\bf k}}

\def\bx{{\bf x}}

\def\CL{{\cal L}}
\def\CN{{\cal N}}
\def\CO{{\cal O}}

\def\tphi{{\tilde \phi}}

\def\mpl{M_{\rm P}}
\def\half{\frac{1}{2}}

\makeatletter
\renewcommand\section{\@startsection {section}{1}{\z@}%
                                 {-3.5ex \@plus -1ex \@minus -.2ex}%
                                   {2.3ex \@plus.2ex}%
                                   {\normalfont\large\bfseries}}
\renewcommand\subsection{\@startsection{subsection}{2}{\z@}%
                                   {-3.25ex\@plus -1ex \@minus -.2ex}%
                                     {1.5ex \@plus .2ex}%
                                     {\normalfont\bfseries}}
\renewcommand\subsubsection{\@startsection{subsubsection}{3}{\z@}%
                                   {-3.25ex\@plus -1ex \@minus -.2ex}%
                                     {1.5ex \@plus .2ex}%
                                     {\normalfont\itshape}}
\makeatother

\newcommand{\Letter}{
\setlength{\textwidth}{16.5cm}
   \setlength{\textheight}{22.6cm}
    \hoffset=-0.5in
\voffset=-2.1cm }

\Letter

\begin{document}
\newcommand{\be}{\begin{equation}}
\newcommand{\ee}{\end{equation}}
\newcommand{\bea}{\begin{eqnarray}}
\newcommand{\eea}{\end{eqnarray}}
\newcommand{\barr}{\begin{array}}
\newcommand{\earr}{\end{array}}

\thispagestyle{empty}
\begin{flushright}
\end{flushright}

\vspace*{0.3in}
\begin{spacing}{1.1}

\begin{center}
{\large \bf Strongly Coupled Inflaton}

\vspace*{0.5in} {Xingang Chen}
\\[.3in]
{\em Center for Theoretical Cosmology, \\
Department of Applied Mathematics and Theoretical Physics, \\
University of Cambridge, Cambridge CB3 0WA, UK } \\[0.3in]
\end{center}

\begin{center}
{\bf
Abstract}
\end{center}
\noindent
We continue to investigate properties of the strongly coupled inflaton in a setup introduced in Ref.~\cite{Chen:2008hz} through the AdS/CFT correspondence. These properties are qualitatively different from those in conventional inflationary models. For example, in slow-roll inflation, the inflaton velocity is not determined by the shape of potential; the fine-tuning problem concerns the dual infrared geometry instead of the potential; the non-Gaussianities such as the local form can naturally become large.

\vfill

\newpage
\setcounter{page}{1}


\newpage

\section{Introduction}
\setcounter{equation}{0}

Inflation models are usually constructed in terms of weakly coupled theories, in which the inflatons and their fluctuations are weakly coupled degrees of freedom in 4D effective field theories and can be studied perturbatively. Nonetheless, strongly coupled theories are an indispensable part of Nature, for example as a building block of the Standard Model. The inflatons may also be the strongly coupled degrees of freedom in 4D effective theories, such as the scalar glueballs in Yang-Mills theory. Properties of such theories are difficult to study in the past. However, the discovery of the AdS/CFT correspondence \cite{Maldacena:1997re,Gubser:1998bc,Witten:1998qj} provides a class of toy models in which strongly coupled theories can be studied perturbatively in higher dimensional dual theories.

In Ref.~\cite{Chen:2008hz}, we started to investigate properties of a 4D strongly coupled inflaton by placing a higher dimensional scalar in AdS$_5$ space deformed by 4D inflationary spacetime. We found that the rolling speed of this scalar in 4D is determined by an eigenvalue, which is analogous to the inflaton mass in conventional slow-roll models. The difference is that, in the conventional models, the inflaton mass is determined by the curvature of inflaton potential; but here this eigenvalue is determined by the infrared (IR) property of the higher dimensional AdS space, which also sensitively depends on the 4D inflationary background. In generic cases, this eigenvalue is of order the Hubble parameter $H$, making the scalar roll down too quickly to become an inflaton. Therefore we saw a problem analogous to the $\eta$-problem in the usual slow-roll inflation models, but manifest in a very different way. The emphasis of Ref.~\cite{Chen:2008hz} is to compare the origin of this $\eta$-problem to that of the $h$-problem \cite{Chen:2008hz} in DBI inflation \cite{Silverstein:2003hf,Chen:2004gc}. They are two completely different inflationary mechanisms. Because the strongly coupled inflaton is interesting enough in its own right, in this paper we investigate it separately in much more details. We will use the same setup, but elaborate it with more model building ingredients. More importantly, we will investigate the following two main issues in the field of inflation model building.

The first is the fine-tuning problem. In conventional slow-roll models, it is well-known that the backreaction of inflationary spacetime makes the inflaton mass generically of order $H$, spoiling the inflation. The identification of the origin of this generic mass term has a special importance -- it allows us to introduce other model-dependent sources or symmetries to solve this problem. It turns out that this mass term can be traced to the canonical inflaton dependence in the Kahler potential in supersymmetrically completed theories \cite{Copeland:1994vg}, or the order one coupling between the Ricci scalar and inflaton. It should be an equally important question for the strongly coupled inflation models whether there exists a generic fine-tuning problem, and if yes where the generic origin is. This is an extension of the study in Ref.~\cite{Chen:2008hz}. As we will see, the effective inflaton mass is not given by the shape of the inflaton potential. Rather it is determined by the IR geometry in the dual higher dimensional theory. In 4D point of view, this is the low energy deformation of the strongly coupled theory at the scale $H$, which also breaks the possibly existed supersymmetry and conformal symmetry.
This is the generic origin of the $\eta$-problem and the place where the effective tuning should take place if we wish to solve this problem.

The second important issue is the density perturbations. This gives the direct observational consequence of inflation and is crucial for the experimental test. It includes the power spectrum (the two-point correlation function) and non-Gaussianities (the higher point correlation functions). The power spectrum is a function of one momentum mode and not very discriminative to different models. Except for possible small corrections due to features, it is plausible that, once a successful inflation model is built, its power spectrum is approximately scale-invariant. This is because a prolonged inflation stretches the inflaton quantum fluctuations with different wavelength in an approximately scale-invariant way. So a more powerful probe is the higher point correlation functions. They have more complicated dependence on momentum modes and contain much more information. However the conventional slow-roll models tend to give non-observable non-Gaussianities. Non-Gaussianities become large only through various special mechanisms \cite{Chen:2010xk}. As we will see, the slow-roll models with strongly coupled inflaton produce much larger non-Gaussianities due to the strong interaction.

Although we have emphasized the point of view that the model to be studied here is the dual of a 4D theory, in which the 10D scalar dilaton with proper boundary condition is dual to the scalar glueball in 4D \cite{Gubser:1998bc,Witten:1998qj}, the higher dimensional picture can also be taken directly and regarded as the model building in warped compactification with non-trivial higher dimensional inflatons.

\section{The setup}

We consider a strongly coupled 4D $\CN=4$ $SU(N)$ super-Yang-Mills theory and couple this sector to 4D gravity. The dual picture is a warped space with ${\rm AdS}_5 \times S^5$ attached to a compact 6D bulk in type IIB string theory. The moduli stabilization of configurations similar to this has been an outstanding problem and actively investigated. We assume this is achieved, for example, in terms of the flux compactification \cite{Douglas:2006es}. The canonical example we have in mind is the GKP-type warped compactification \cite{Giddings:2001yu}. The dynamical degrees of freedom in this theory can be most easily studied on the AdS side, so we will work on this side of the duality. We will ignore the angular part and focus on the most important 5D metric,
\bea
ds^2 = G_{(5)MN} dX^M dX^N= h(r)^2 (-dt^2 + a(t)^2 d\bx^2 ) + h(r)^{-2} dr^2 ~,
\label{5Dmetric}
\eea
where $h(r)$ is the warp factor with length scale $R$, and $a$ is the inflationary scale factor. This metric is asymptotically AdS as the energy scale $\gg H$ and $h(r) \to r/R$. The warp factor $h(r)$ smoothly becomes $\CO(1)$ at UV ($r\sim R$) where the warped throat is attached to a compactified bulk of size $L$. At the IR end of the warped space, we will impose a sharp cutoff at $r=r_c$. The value of $r_c$ will be discussed in the next section.

Consider a 5D scalar field $\phi(t,\bx,r)$ with mass $M_B$ in this background. To introduce an inflationary potential, we consider a 4D scalar field $\tphi(t,\bx)$ on a spacetime-filling brane at $r=r_b$. The $\tphi$ can be induced by $\phi$ through couplings. We put this brane near the UV end of the warped space, $r_b\sim R$. The induction strength is parameterized by $1/\alpha$, $\phi(r_b,t,\bx) = \alpha \tphi(t,\bx)$.
The inflation potential is given by a 4D potential $V_b(\tphi)$,
\bea
V_b = V_{b0} + \half m_b^2 \tphi^2 + \cdots ~.
\label{Pot_Vb}
\eea
We assume that the inflation ends when $\tphi$ reaches a critical value $\tphi_{\rm end}$, which for example can be engineered by adding a tachyonic instability.

The action is
\bea
S &=& \int d^4x dr \sqrt{-G_{(5)}}
\left[ \frac{M_{(5)}^3}{2} R_{(5)} + \CL_{\rm matter}(\Phi) \right]
\nonumber \\
&+&
\int d^4x dr \sqrt{-G_{(5)}}
\left[ - \half G_{(5)}^{MN} \partial_M \phi \partial_N \phi - \half M_B^2 \phi^2 \right]
\nonumber \\
&+&
\int d^4x \sqrt{-G_{(4)}}
\left[ - \half G_{(4)}^{\mu\nu} \partial_\mu \tphi \partial_\nu \tphi - V_b(\tphi) \right]
~.
\label{5DAction}
\eea
The $\CL_{\rm matter}$ represents the Lagrangian of the moduli-stabilizing fields which we collectively denote as $\Phi$.
The induced 4D metric is $G_{(4)\mu\nu} = G_{(5)\mu\nu}|_{r_b}$. The inflation is driven by the potential energy $V_{b0}$, $H^2 = V_{b0}/(3 \mpl^2)$, where the Planck mass is obtained from $M_{(5)}$ by integrating over the extra dimension of size $L \gtrsim R$, $\mpl^2 \sim M_{(5)}^3 L$.
The 5D scalar $\phi$ has wave-functions in both the spacetime and $r$-direction. We decompose it as
\bea
\phi(r,t,\bx) = \sum_n f_n(r) \varphi_n(t,\bx) ~.
\eea
The linear equation of motion for $\phi$ can be separated into the following two equations for each $n$,
\bea
\frac{1}{h} \partial_r (h^5 \partial_r f_n) - h^2 M_B^2 f_n + \lambda_n^2 f_n + (\lambda_n^2 - h^2 m_b^2) \frac{h}{\alpha^2} f_n \delta(r-r_b) &=& 0 ~,
\label{fEOM}
\\
\ddot \varphi_n + 3H \dot\varphi_n - \frac{1}{a^2} \nabla^2 \varphi_n + \lambda_n^2 \varphi_n &=& 0 ~.
\label{phiEOM}
\eea
Equation (\ref{fEOM}) is an eigenvalue problem for $\lambda_n$, and the eigenfunctions $f_n$ are normalized as
\bea
\frac{h^2}{\alpha^2} f_m f_n |_{r_b} + \int_{r_c}^L dr h f_m f_n = \delta_{mn} ~.
\label{fNormCond}
\eea
The boundary condition for $f_n$ at $r_c$ is $\partial_r f_n |_{r_c}=0$ or $f_n (r_c) =0$.
After we integrate the wave-function $f_n(r)$ over $r$,
the kinetic term for each 4D scalar field $\varphi_n$ is canonical.
For the zero-mode background, the slow-roll of $\varphi_n$ is given by
\bea
3 H \dot \varphi_{n0} + \lambda_n^2 \varphi_{n0} = 0.
\label{varphiEOM_sr}
\eea
The rolling velocity is determined by the eigenvalue $\lambda_n$, which plays the role of the effective 4D scalar mass. This is indirectly related to the shape of the potentials $m_b$ or $M_B$ through (\ref{fEOM}). As we will see in more details later, for the strongly coupled inflaton the most important properties of $\lambda_n$ are not determined by these shapes.

Before proceed, we would like to clarify which aspects of the model are more model-dependent or model-independent. The first line of (\ref{5DAction}) provides the stabilized compactification involving model-dependent details. The fields $\phi$ and $\tphi$ are probes of this compactification. Namely, the inflationary Hubble energy they introduce is much lower than the mass of the moduli fields $\Phi$, except for the IR end of the AdS space. The third line of (\ref{5DAction}) is introduced to raise the 4D vacuum energy necessary for inflation, which is also model-dependent. The effective relation between $\phi$ and $\tphi$ we used may be realized microscopically through an interaction action $\int d^4x \sqrt{-G_{(4)}} C \tphi \phi|_{r=r_b}$. The $\phi$ field evolution, determined by (\ref{fEOM}) and (\ref{phiEOM}), becomes an external force for the evolution of $\tphi(t)$,
\bea
\ddot \tphi + 3H \dot \tphi + h_b^2 m_b^2 \tphi = C h_b^2 \phi|_{r_b} ~.
\label{eom_tildephi}
\eea
The solution $\tphi(t)$ is a summation of the non-homogeneous part, $\phi/\alpha|_{r_b}$, and the source-free homogeneous part, where the induction strength parameter we used is now given by $\alpha = (-\lambda^2 +h_b^2m_b^2)/(Ch_b^2)$ and $h_b$ denotes $h|_{r_b}$. Here we have assumed that the most important $\lambda_n$ are all similar $\sim \lambda$, as we will see later. For the potential (\ref{Pot_Vb}) with a positive and naturally large mass-squared, $h_b^2 m_b^2 \sim H^2$, the homogeneous part of $\tphi$ will quickly decrease and get red-shifted. More complicated couplings such as $\sim \tphi \phi^n$ will lead to more complicated induction relations such as $\tphi \propto \phi^n$. However, most of the properties that we will study in the rest of the paper is much more model-independent. Namely, we will investigate the properties of a test 5D scalar field in the AdS$_5$ space deformed by the 4D inflationary spacetime, by studying the second line of (\ref{5DAction}) (and its non-linear extension) in the background of (\ref{5Dmetric}). In addition, this test scalar field may be used either as the inflaton, as we do in this paper, or as the isocurvaton or curvaton.

\section{The fine-tuning problem}

The inflationary spacetime generically imposes a cutoff on the AdS-like warp factor at
\bea
h_c \sim HR ~.
\label{hc_Generic}
\eea
The origin can be seen by considering the left hand side of the following Einstein equation,
\bea
(\partial_r h)^2 - H^2 h^{-2} = 1/R^2 + {\rm other~source~terms} ~.
\eea
Due to the inflationary backreaction term $H^2 h^{-2}$, the naive AdS-like geometry $h=r/R$ generically has to be deformed at (\ref{hc_Generic}), unless tuned by other source terms. For simplicity, we will approximate the deformed warp factor as $h=r/R$ with a sharp cutoff at $h_c=HR$.

We first consider the eigenvalue problem (\ref{fEOM}) without the delta-function term. The general solution is a linear combination of two components,
\bea
f_n(r)= c_1 \frac{R^2}{r^2} J_\nu ( \lambda_n R^2/r ) +
c_2 \frac{R^2}{r^2} Y_\nu ( \lambda_n R^2/r ) ~,
\eea
where $J_\nu$ and $Y_\nu$ are the Bessel function of the first and second kind, respectively, and $\nu = \sqrt{4+M_B^2 R^2}$. As an example, we consider the case $M_B R \ll 1$, so $\nu \approx 2$.
The two components have different UV behavior in the decompactification limit, which in the normalization condition (\ref{fNormCond}) is $L\to \infty$ with $h=r/R$.
The first component is normalizable and it peaks at IR. The second is not normalizable, and it is nearly a constant for $r/R \gg \lambda_n R$. In the IR, in order to have non-singular behavior at $h=h_c$, we require
\bea
\partial_r f_n|_{r_c}=0 ~.
\label{BCond}
\eea

If we only consider the normalizable component, it is easy to see that the boundary condition (\ref{BCond}) requires the eigenvalues $\lambda_n$ be positive, discrete and of order $\CO(H)$: $\lambda_n = 3.83H, 7.02H, \dots$. Therefore the effective mass for the scalars $\varphi_n$ is of order $H$, too large to support the slow-roll in (\ref{varphiEOM_sr}). Following the conventional terminology, we call this problem the ``$\eta$-problem" despite of the fact that the $\lambda_n$ has a different nature from the conventional $\eta$.

Because the extra dimension is compactified, the upper bound in the normalization condition (\ref{fNormCond}) is finite. Taking this into account, the second component also becomes normalizable. By mixing this component to the first one, we can move each of the above discrete eigenvalues very close to zero.
In order to do this, by expanding the Bessel functions in the small $\lambda_n R^2/r$ limit, we can see that we need $c_1/c_2 \sim h_c^2/(\lambda_n^2 R^2)$ to satisfy (\ref{BCond}). For example, if $c_1/c_2 = 100$, we get $\lambda_n=0.11H, \dots$.
In this case the IR deformation of the warped space does not contribute to the $\eta$-problem, and the wave-function $f_n$ is mostly dominated by the second component.

The physical interpretation of the above two cases is as follows. As mentioned, the wave-function of the inflaton in the first component is concentrated near the IR end of the warped space. So in terms of the dual 4D field theory, this inflaton is the strongly coupled scalar glueball. In this case, we see that the inflation-deformed IR geometry presents an $\eta$-problem.
The wave-function in the second component smears out in the UV and bulk. With suitable mixing with the first component, this inflaton is the conventional weakly coupled component since the AdS space does not play any role when we integrate out the extra dimension. The $\eta$-problem should arise due to the physics in the bulk as in the weakly coupled theory.\footnote{For example, in the bulk $h\sim 1$, so from (\ref{fEOM}) we have $f_n \sim {\rm constant}$ and $\lambda_n^2 \sim M_B^2$. In the case where the wave-function of $\phi$ is dominated by the bulk part, $M_B^2$ is naturally of order $H^2$.}
In this paper we are interested in the former case, so we approximate
\bea
f_n(r) = c \frac{R^2}{r^2} J_2 ( \lambda_n R^2/r ) ~.
\label{f_Wavefun}
\eea

To see how the shape of potential $V_b$ affects the above conclusion, we look at the term involving the delta-function at $r_b$ in (\ref{fEOM}). This term gives the junction condition for $f_n$ across $r_b$,
\bea
\partial_r f_n|_-^+ =  \frac{-\lambda_n^2 + h_b^2 m_b^2}{h_b^3 \alpha^2} f_n ~,
\quad
f_n|_-^+ = 0 ~.
\eea
Since $r_b$ is at the UV side of the warped space, the situation for $r<r_b$ is exactly the same as what we have discussed. Therefore, for the strongly coupled inflaton, the $\CO(H)$ mass gap for $\lambda_n$ is not affected by $m_b$.\footnote{Recall that $h_b m_b$ is the mass for the homogeneous component in the solution of the differential equation (\ref{eom_tildephi}). This component corresponds to a usual 4D weakly coupled field, on the brane. As we described, the non-homogeneous component of $\tilde\phi$ in (\ref{eom_tildephi}) is driven by the strongly coupled inflaton $\phi$.}
The value of $m_b$, along with other parameters such as $\alpha$, determines the wave-function in the bulk $r>r_b$. This part of the wave-function is highly model dependent. For example, the jump in $\partial_r f_n$ can be negligible if $\alpha^2 \gg |-\lambda_n^2 + h_b^2 m_b^2| R/h_b^2$. We will concentrate on this case in the rest of the paper, since it emphasizes the properties of the strongly coupled inflaton. The jump can also be significant so the wave-function have peaks in both the IR end and the bulk, in which case the bulk part dilutes the properties of the strongly coupled inflaton that we will study later.

To summarize, the IR cutoff of the AdS geometry due to the inflationary backreaction is generically at $h_c \sim HR$. This deformation affects the wave-function of the strongly coupled inflaton at IR so that the 4D effective mass of the inflaton becomes of order $H$. This is the $\eta$-problem, but not determined by the inflationary potential. In order to solve this problem, we need to adjust the other source terms for the warped geometry, so that the cutoff is pushed to a much smaller value, $\bar h_c \sim \CO(10^{-1}) HR$. Replacing $h_c$ with $\bar h_c$ in the above analyses gives the discrete mass gap of order $\lambda_n \sim \bar h_c/R \sim \CO(10^{-1}) H$. How this can be achieved in the actual model-building is the fine-tuning problem, which we expect to be a very rich and interesting subject. We leave it for future study. In the rest of this paper, we will assume that such a fine-tuning has been achieved and go on to investigate the observational consequences.

In realistic scenarios, the strongly coupled scalars and the conventional weakly coupled scalars coexist and they require different types of fine-tuning for them to become inflatons. If the dual geometries for the former have been fine-tuned as above, but the potentials for the latter have not, the weakly coupled scalars will just roll down quickly, leaving only the strongly coupled ones as the successful inflatons, which is the case of interest here; and vice versa.

\section{Power spectrum}

We have seen that there can be many inflatons with discrete eigenvalues $\lambda_n$ of order $\bar h_c /R$ and higher. A precise computation should take into account all of them. It is plausible that the more important ones are those with lower eigenvalues because they roll down more slowly. So in this paper, we will only consider one eigenfunction with the lowest effective mass $\lambda \sim \bar h_c/R$.

The inflation ends when the 4D scalar $\tphi$ reaches a critical value. The space-dependent time-delay is $\delta t = -\delta \tphi/\dot \tphi = -\delta \varphi/\dot \varphi$, where $\dot\varphi$ is determined by (\ref{varphiEOM_sr}).
So the map between the 4D observable, namely the density perturbation $\zeta$, and the 5D field $\phi$ is $\zeta = H\delta t= -H\delta\phi/\dot\phi |_{r_b}= -H\delta\varphi/\dot\varphi$.
Since $\varphi$ has the canonical kinetic term under the normalization (\ref{fNormCond}), it has the usual mode function decomposition
\bea
\delta\varphi &=& u(\bk,\tau) a_\bk + u^*(-\bk,\tau) a_{-\bk}^* ~,
\label{delt_avarphi_Decomp}
\\
u(\bk,\tau) &=& \frac{H}{\sqrt{2k^3}} (1+ik\tau) e^{-ik\tau} ~,
\eea
where $\tau = \int dt/a$ is the conformal time and $a_\bk$ satisfies the usual canonical commutation relation. This corresponds to $\delta\varphi=H/(2\pi)$ in the above formula for $\delta t$.
The density perturbation is then
\bea
\zeta = H\delta t = -\frac{3H^3}{2\pi \lambda^2 \varphi} ~.
\eea
The spectral index is
\bea
n_s-1 = 2 \frac{d\ln \zeta}{H dt}
= -6\epsilon + \frac{2 \lambda^2}{3 H^2} ~,
\eea
where $\epsilon \equiv -\dot H/H^2$ is as usual, but the conventional $\eta_V=M_p V''/V$ has been replaced by $\lambda^2/(3H^2)$. The condition $\bar h_c \sim \CO(10^{-1}) RH$ is necessary to ensure an approximately scale-invariant density perturbations, as well as to provide a prolonged inflation.

An approximate scale-invariant power spectrum can be easily produced by many other inflation models, hence is not observationally very distinctive. We now turn to the three-point scalar perturbation, i.e.~the bispectrum.

\section{Bispectrum}

We consider two types of non-linearities. One is in the bulk, another is on the localized brane. Non-linearity is intrinsic to the model due to gravity. However, a full perturbation theory including both gravity and matter is beyond the scope of the current paper. To show the most important qualitative properties, we will proceed by introducing some non-linear terms in the scalar sector and studying the three-point perturbations. Our emphasis will be to compare the conventional weakly coupled case with the strongly coupled one.

The first example is the bulk non-linearity. We introduce the following action term for the 5D field $\phi$,
\bea
- \kappa \int d^4x dr \sqrt{-G_{(5)}}  G_{(5)}^{MN} \phi \partial_M \phi \partial_N \phi ~.
\label{BulkNonlinear}
\eea
This term introduces a cubic interaction for $\delta\phi$ with the same structure.
At this level, this is just one particular example of cubic terms.\footnote{In usual slow-roll models, it is not enough to fine-tune the slow-roll conditions at one field value, the conditions have to be maintained and tuned at all field values the inflaton take during inflation. Similarly, here we assume that the corrections introduced by (\ref{BulkNonlinear}) on the zero-mode inflaton evolution are tuned away by adjusting the geometry along the inflaton trajectory. So we will only consider the fluctuations.} We are allowed to add this term as long as it is a small perturbation, so we choose a very small coupling, $\kappa = \CO(\epsilon) P_\zeta^{1/2} H^{-1} R^{1/2}$, where $\epsilon \sim \CO(10^{-2})$ and $P_\zeta \sim \CO(10^{-10})$ is the power spectrum. As we will see shortly, this example is chosen because, in the usual weakly coupled slow-roll inflation, its structure and size are qualitatively similar to those arising from the non-linearity of gravity.

The three-point function of $\delta\varphi$ contributed by (\ref{BulkNonlinear}) can be computed with (\ref{delt_avarphi_Decomp}) following the standard procedure \cite{Chen:2010xk}, and that of $\zeta$ can be obtained using the relation $\zeta= -H\delta\varphi/\dot\varphi$. The resulting bispectrum has a local shape, which in terms of the shape function $S$ \cite{Chen:2010xk} is
\bea
S(k_1,k_2,k_3) = \frac{3}{10} f_{NL} \frac{\sum_i k_i^3}{k_1 k_2 k_3} ~,
\eea
where
\bea
f_{NL} = \frac{5}{24\pi} H P_\zeta^{-1/2} \kappa \int_{r_c}^L dr h f^3 ~.
\label{fNL_Bulk}
\eea

If the extra dimension is flat with the size $L\sim R$, we recover the normal weakly coupled inflaton after integrating out this dimension. In this case, $h$ is a constant. Using (\ref{fNormCond}) to determine the normalization constant for $f$ and evaluating the remaining integral in (\ref{fNL_Bulk}), we get
\bea
f_{NL}^{\rm weak} \sim \CO(\epsilon) ~.
\eea
For the AdS geometry and the strongly coupled inflaton (\ref{f_Wavefun}), the same procedure leads to
\bea
f_{NL}^{\rm strong} \sim \CO(\epsilon) \bar h_c^{-1} ~.
\eea
It has a large enhancement factor $\bar h_c^{-1} \gg 1$.
The difference between the two results is caused by the different wave-function distribution in the extra dimension.
From (\ref{fNL_Bulk}), we can see that,
due to the normalization [the 2nd term in (\ref{fNormCond})], more localized wave-function leads to larger interaction. The wave-function of the strongly coupled inflaton concentrates near the scale $r/R\sim \bar h_c$, and this is where the large non-Gaussianity is generated. We also note that the effect of fine-tuning is opposite for the two cases. For conventional slow-roll models, making potential flatter reduces the non-linearity and decreases the non-Gaussianity; while for strongly coupled slow-roll models, making the warp factor smaller enhances the bulk interactions and increases the non-Gaussianity.

The second example is the non-linearity localized in the inflationary potential at UV. We consider the following cubic interaction term from the potential $V_b$,
\bea
- \int d^4 x \sqrt{-G_{(4)}} \frac{1}{6} V_b''' \delta\tphi^3 ~.
\eea
This term gives a bispectrum shape close to the local form\footnote{In this bispectrum integral we need to regulate the unphysical IR divergence by taking the upper limit to be the time of the horizon exit for some momentum modes (e.g.~$k_1+k_2+k_3$).} with
\bea
f_{NL}^{\rm strong} \sim \frac{h_b^3 f_b^3}{\alpha^3} \frac{V_b'''}{H} P_\zeta^{-1/2} ~.
\label{LocfNLstrong}
\eea
In the familiar weakly coupled slow-roll model with a potential $V$, the same type of cubic interaction gives
\bea
f_{NL}^{\rm weak} \sim \frac{V'''}{H} P_\zeta^{-1/2} ~.
\label{LocfNLweak}
\eea
Unlike the previous bulk non-linearity case, here the two expressions are similar. However, in (\ref{LocfNLweak}), $V'''$ has to be tuned to be of order $\CO(\epsilon^2) P_\zeta^{1/2} H$ to maintain the slow-roll condition; while here $V_b'''$ is not directly related to the slow-roll condition, so can be very large naturally. Therefore (\ref{LocfNLstrong}) can become large (even with $h_b \lesssim 1$ and $f_b/\alpha \ll 1$).

To summarize, we have seen two different mechanisms that can enhance the inflaton interactions due to the non-trivial wave-function in the AdS direction. First, the wave-function peaks in the IR and increases bulk interactions there. Second, away from IR, local interactions do not affect the effective inflaton mass, and hence are not constrained to be small. In conventional models, large local non-Gaussianity can only be generated by superhorizon classical evolution \cite{Lyth:2005fi}, or by isocurvature modes quantum-mechanically at horizon scale and then transferred to the curvature perturbation \cite{Chen:2009zp}. Here we see that the same type of local non-Gaussianity present in the simplest single field slow-roll inflation models, which is generated quantum-mechanically at the horizon scale and directly for the curvature mode, can naturally become large due to the strong interaction.

\section{Discussions}

We make some comparison with several well-known inflationary models involving the AdS space. In the higher dimensional theory, these models describe branes moving in the warped compactification \cite{HenryTye:2006uv}.
In the 4D field theory, these brane positions are weakly coupled scalar fields. For example, the KKLMMT model \cite{Kachru:2003sx} describes a brane slowly moving in a warped throat. Once the potential is given, the 4D effective field theory is the usual slow-roll inflation. In DBI brane inflation models \cite{Silverstein:2003hf,Chen:2004gc}, this brane moves with a velocity approaching the local speed of light. The inflaton in the 4D effective theory is still weakly coupled but becomes relativistic, with non-canonical kinetic terms.\footnote{In \cite{Evans:2010tf}, as an interesting extension of brane inflation, the D3 probe brane is replaced by D7 probe brane, so the rolling of a point in the extra dimensions becomes the rolling of a line. If we use the moduli space approximation, the 4D effective theory should become the weakly coupled single or multi-field slow-roll/DBI inflation model.} In other words, the interactions in these theories can be studied perturbatively, at least in the low energy limit, using the dynamical degrees of freedom explicitly present in the 4D effective Lagrangian; while for the strongly coupled inflaton studied here, this is impossible and one has to use the dual higher dimensional theory.

There are many outstanding issues remain to be investigated. For example, it is important to build the models more explicitly and in more varieties, especially to provide examples where the fine-tuning is realized. Perturbation theory involving both the matter and gravity sector needs to be investigated. The IR energy scale is at least $\CO(10)$ times larger than the local KK scale, so effects from multiple particles and even strings can be interesting and may provide more stringy predictions.

\medskip
\section*{Acknowledgments}

I am supported by the Stephen Hawking advanced fellowship.

\end{spacing}

\end{document}